\RequirePackage{lineno}
\documentclass[aps,prl,twocolumn,showpacs,amsmath,preprintnumbers,amssymb,groupedaddress,superscriptaddress]{revtex4-1}
\usepackage{graphicx}
\usepackage{dcolumn}
\usepackage{amsmath}

\def\sNN{\mbox{$\sqrt{s_{_{NN}}}$}}  
\newcommand{ \be }{\begin{equation}}      
\newcommand{ \ee }{\end{equatiton}}      
\newcommand{ \bea }{\begin{eqnarray}}      
\newcommand{ \eea }{\end{eqnarray}}

\usepackage{color}

\begin{document}


\title{
\begin{flushright} 
{\small \sl version 07,  \today \\ 
 }      
\end{flushright} 
Measurement of $J/\psi$ Azimuthal Anisotropy in Au+Au Collisions at  \sNN = 200 GeV}



\affiliation{AGH University of Science and Technology, Cracow, Poland}
\affiliation{Argonne National Laboratory, Argonne, Illinois 60439, USA}
\affiliation{University of Birmingham, Birmingham, United Kingdom}
\affiliation{Brookhaven National Laboratory, Upton, New York 11973, USA}
\affiliation{University of California, Berkeley, California 94720, USA}
\affiliation{University of California, Davis, California 95616, USA}
\affiliation{University of California, Los Angeles, California 90095, USA}
\affiliation{Universidade Estadual de Campinas, Sao Paulo, Brazil}
\affiliation{Central China Normal University (HZNU), Wuhan 430079, China}
\affiliation{University of Illinois at Chicago, Chicago, Illinois 60607, USA}
\affiliation{Cracow University of Technology, Cracow, Poland}
\affiliation{Creighton University, Omaha, Nebraska 68178, USA}
\affiliation{Czech Technical University in Prague, FNSPE, Prague, 115 19, Czech Republic}
\affiliation{Nuclear Physics Institute AS CR, 250 68 \v{R}e\v{z}/Prague, Czech Republic}
\affiliation{University of Frankfurt, Frankfurt, Germany}
\affiliation{Institute of Physics, Bhubaneswar 751005, India}
\affiliation{Indian Institute of Technology, Mumbai, India}
\affiliation{Indiana University, Bloomington, Indiana 47408, USA}
\affiliation{Alikhanov Institute for Theoretical and Experimental Physics, Moscow, Russia}
\affiliation{University of Jammu, Jammu 180001, India}
\affiliation{Joint Institute for Nuclear Research, Dubna, 141 980, Russia}
\affiliation{Kent State University, Kent, Ohio 44242, USA}
\affiliation{University of Kentucky, Lexington, Kentucky, 40506-0055, USA}
\affiliation{Institute of Modern Physics, Lanzhou, China}
\affiliation{Lawrence Berkeley National Laboratory, Berkeley, California 94720, USA}
\affiliation{Massachusetts Institute of Technology, Cambridge, MA 02139-4307, USA}
\affiliation{Max-Planck-Institut f\"ur Physik, Munich, Germany}
\affiliation{Michigan State University, East Lansing, Michigan 48824, USA}
\affiliation{Moscow Engineering Physics Institute, Moscow Russia}
\affiliation{National Institute of Science and Education and Research, Bhubaneswar 751005, India}
\affiliation{Ohio State University, Columbus, Ohio 43210, USA}
\affiliation{Old Dominion University, Norfolk, VA, 23529, USA}
\affiliation{Institute of Nuclear Physics PAN, Cracow, Poland}
\affiliation{Panjab University, Chandigarh 160014, India}
\affiliation{Pennsylvania State University, University Park, Pennsylvania 16802, USA}
\affiliation{Institute of High Energy Physics, Protvino, Russia}
\affiliation{Purdue University, West Lafayette, Indiana 47907, USA}
\affiliation{Pusan National University, Pusan, Republic of Korea}
\affiliation{University of Rajasthan, Jaipur 302004, India}
\affiliation{Rice University, Houston, Texas 77251, USA}
\affiliation{Universidade de Sao Paulo, Sao Paulo, Brazil}
\affiliation{University of Science \& Technology of China, Hefei 230026, China}
\affiliation{Shandong University, Jinan, Shandong 250100, China}
\affiliation{Shanghai Institute of Applied Physics, Shanghai 201800, China}
\affiliation{SUBATECH, Nantes, France}
\affiliation{Temple University, Philadelphia, Pennsylvania, 19122}
\affiliation{Texas A\&M University, College Station, Texas 77843, USA}
\affiliation{University of Texas, Austin, Texas 78712, USA}
\affiliation{University of Houston, Houston, TX, 77204, USA}
\affiliation{Tsinghua University, Beijing 100084, China}
\affiliation{United States Naval Academy, Annapolis, MD 21402, USA}
\affiliation{Valparaiso University, Valparaiso, Indiana 46383, USA}
\affiliation{Variable Energy Cyclotron Centre, Kolkata 700064, India}
\affiliation{Warsaw University of Technology, Warsaw, Poland}
\affiliation{University of Washington, Seattle, Washington 98195, USA}
\affiliation{Wayne State University, Detroit, Michigan 48201, USA}
\affiliation{Yale University, New Haven, Connecticut 06520, USA}
\affiliation{University of Zagreb, Zagreb, HR-10002, Croatia}

\author{L.~Adamczyk}\affiliation{AGH University of Science and Technology, Cracow, Poland}
\author{J.~K.~Adkins}\affiliation{University of Kentucky, Lexington, Kentucky, 40506-0055, USA}
\author{G.~Agakishiev}\affiliation{Joint Institute for Nuclear Research, Dubna, 141 980, Russia}
\author{M.~M.~Aggarwal}\affiliation{Panjab University, Chandigarh 160014, India}
\author{Z.~Ahammed}\affiliation{Variable Energy Cyclotron Centre, Kolkata 700064, India}
\author{I.~Alekseev}\affiliation{Alikhanov Institute for Theoretical and Experimental Physics, Moscow, Russia}
\author{J.~Alford}\affiliation{Kent State University, Kent, Ohio 44242, USA}
\author{C.~D.~Anson}\affiliation{Ohio State University, Columbus, Ohio 43210, USA}
\author{A.~Aparin}\affiliation{Joint Institute for Nuclear Research, Dubna, 141 980, Russia}
\author{D.~Arkhipkin}\affiliation{Brookhaven National Laboratory, Upton, New York 11973, USA}
% ...


\collaboration{STAR Collaboration} \noaffiliation

\date{\today}

\begin{abstract}
The measurement of $J/\psi$ azimuthal anisotropy is presented as a function of transverse momentum for different centralities in Au+Au collisions at  \sNN = 200 GeV. The measured $J/\psi$ elliptic flow is consistent with zero within errors for transverse momentum between 2 and 10 GeV/$c$. Our measurement suggests that $J/\psi$ with relatively large transverse momentum are not dominantly produced by coalescence from thermalized charm quarks, when comparing to model calculations.
\end{abstract}


\maketitle



Quantum chromodynamics (QCD) predicts a quark-gluon plasma (QGP) phase at extremely high temperature and/or density, consisting of deconfined quarks and gluons. Over the past twenty years, heavy quarkonia production in hot and dense nuclear matter has been a topic attracting growing interest. In relativistic heavy-ion collisions the $c\bar{c}$ bound state is subject to dissociation due to the color screening effect in the deconfined medium. As a consequence, the production of $J/\psi$ is expected to be suppressed compared to proton+proton ($p+p$) collisions scaled by number of binary collisions, and such suppression has been proposed as a signature of QGP formation \cite{colorscreen}. However, the $J/\psi$ suppression observed in experiments \cite{Abreu:2000xe,PhysRevLett.98.232301,Adare:2007gn,Zebo:2012,Alice:2012} can also be affected by additional cold \cite{Johnson:2000ph,Guzey:2004zp} and hot \cite{Baier:2000mf,Gavin:1996fe,Thews05,Thews06,Andronic06,Capella07} nuclear effects. In particular the recombination of $J/\psi$ from a thermalized charm quark and its antiquark \cite{Thews05,Thews06,Andronic06,Capella07} has not been unambiguously established experimentally at the top RHIC energy. By measuring $J/\psi$ azimuthal anisotropy, especially its second Fourier coefficient $v_{2}$ (elliptic flow), one may infer the relative contribution of $J/\psi$ from direct pQCD processes and from recombination. $J/\psi$ produced from direct pQCD processes, which do not have initial collective motion, should have little azimuthal preference. In non-central collisions, the produced $J/\psi$ will then gain limited azimuthal anisotropy from azimuthally different absorption due to the different path lengths in azimuth. On the other hand, $J/\psi$ produced from recombination of thermalized charm quarks will inherit the flow of charm quarks, exhibiting considerable flow.

Many models that describe the experimental results of heavy-ion collisions depend on the assumption that light flavor quarks in the medium reach thermalization on a short timescale ($\sim$ 0.5 fm/$c$) \cite{Kolb:2003dz,Huovinen:2006jp}. However, this rapid full thermalization has not been directly certified. The flow pattern of heavy quarks provides a unique tool to test the thermalization. With much larger mass than that of light quarks, heavy quarks are more resistant to having their velocity changed, and are thus expected to thermalize much more slowly than light partons. If charm quarks are observed to have sizable collective motion, then light partons, which dominate the medium, should be fully thermalized. The charm quark flow can be measured through open \cite{PHENIXEV2} and closed charm particles. The $J/\psi$ is the most prominent for experiment among the latter. However, because the $J/\psi$ production mechanism is not well understood, there is significant uncertainty associated with this probe, since only $J/\psi$ from recombination of charm quarks inherit their flow. A detailed comparison between experimental measurements and models on $J/\psi$ $v_{2}$ vs. transverse momentum ($p_{T}$) and centrality, in addition to nuclear modification factor, will shed light on the $J/\psi$ production mechanism and charm quark flow.

This analysis benefits from a large amount of data taken during the RHIC \cite{RHIC} $\sqrt{s_{NN}}$ = 200 GeV Au+Au run in the year 2010 by the new data acquisition system of STAR \cite{STAR}, capable of an event rate up to 1 kHz. In addition, the newly installed Time Of Flight (TOF) detector \cite{TOF} allows STAR to improve electron identification, and background electrons from photon conversion are reduced by one order of magnitude due to less material around the center of the detector setup. The data presented consist of 360 million minimum bias (MB) events triggered by the coincidence of two Vertex Position Detectors \cite{VPD}, 270 million central events triggered by a large hit multiplicity in the TOF detector \cite{TOF}, and a set of high tower events triggered by signals in the towers of Barrel Electromagnetic Calorimeter (BEMC) \cite{BEMC} exceeding certain thresholds (2.6, 3.5, 4.2, and 5.9 GeV). The high tower sample is equivalent to approximately 7 billion MB events for $J/\psi$ production in the high-$p_{T}$ region. In addition, in order to cope with the large data volume coming from collisions at high luminosity, a High Level Trigger (HLT) was implemented to reconstruct charged tracks online, select events with $J/\psi$ candidates and tag them for fast analysis. There are 16 million $J/\psi$ enriched events selected by the HLT.

The $J/\psi$ were reconstructed through the $J/\psi \rightarrow e^+e^-$ channel, which has a branching ratio of 5.9 \%. The daughter tracks of the $J/\psi$ were required to have more than 20 hits in the Time Projection Chamber (TPC) \cite{TPC}, and a distance of closest approach less than 1 cm from the primary vertex. Low momentum electrons and positrons can be separated from hadrons by selecting on the inverse velocity ($0.97 < 1/\beta < 1.03$), which is calculated from the time-of-flight measured by the TOF detector \cite{TOF} and the path length measured by the TPC. At large momentum ($p > 1.5$ GeV/$c$), with the energy measured by towers from the BEMC \cite{BEMC}, a cut of the momentum to energy ratio ($0.3 < p/E < 1.5 $) was applied to select electrons and positrons. The electrons and positrons were then identified by their specific energy loss ($\langle dE/dx \rangle$) inside the TPC. More than 15 TPC hits were required to calculate $\langle dE/dx \rangle$. The $\langle dE/dx \rangle$ cut is asymmetric around the expected value for electrons, because the lower side is where the hadron $\langle dE/dx \rangle$ lies. It also varies according to whether the candidate track passes the $1/\beta$ and/or $p/E$ cut to optimize efficiency and purity. The combination of cuts on $1/\beta$, $p/E$ and $\langle dE/dx \rangle$ enables electron/positron identification in a wide momentum range. Our measured $J/\psi$ particles cover the rapidity range $-1<y<1$, favoring $J/\psi$ near $y=0$ because of detection efficiency variation due to acceptance and decay kinematics. A total of just over 13000 $J/\psi$ were reconstructed in the entire $p_{T}$ range of $0-10$ GeV/$c$.

\begin{figure}
\includegraphics[width=1.\columnwidth]{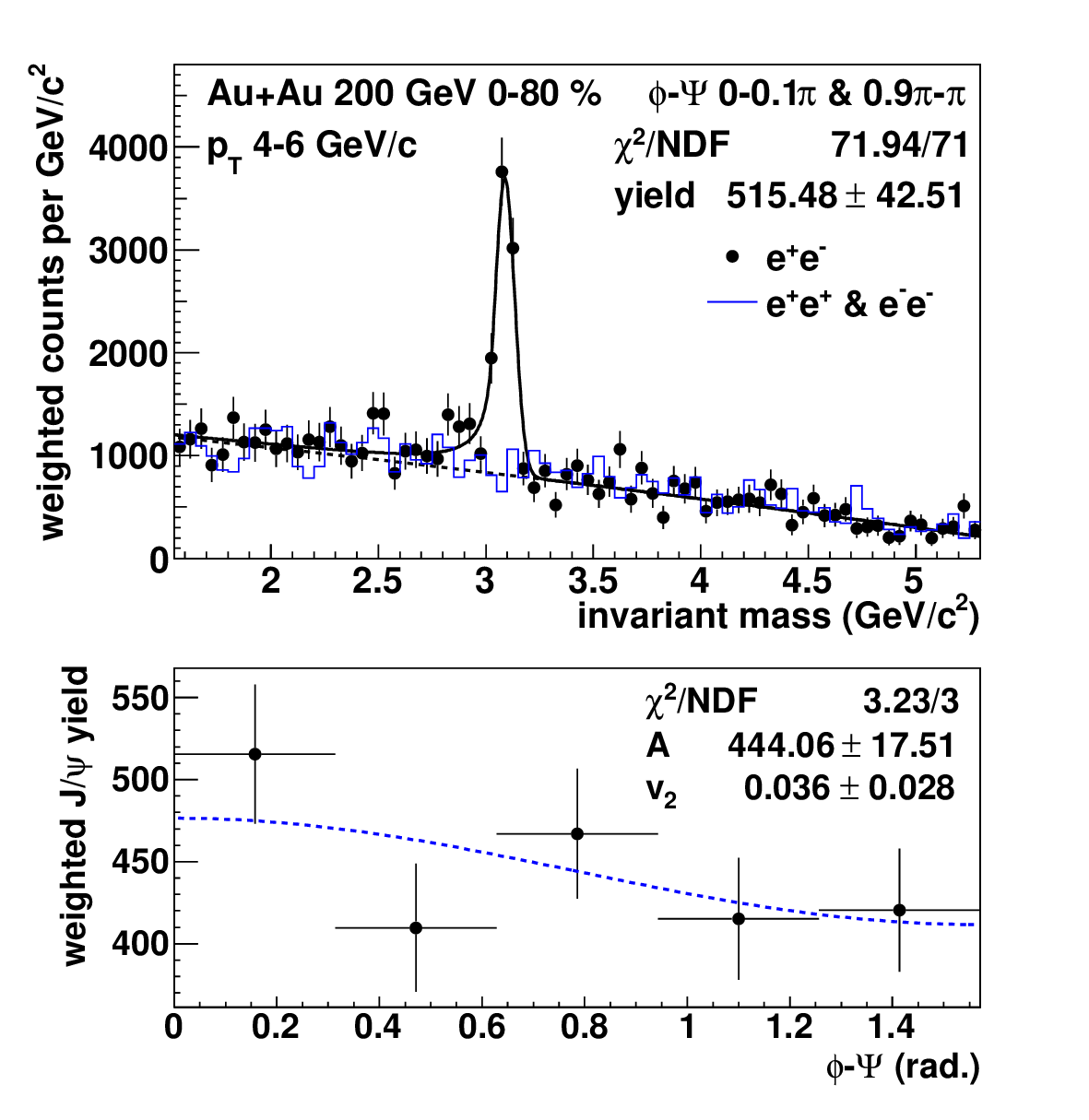}%
\caption{(color online) Top: $1/R$ weighted invariant mass spectrum of electron/positron pairs for $\phi$-$\Psi$ in $0-0.1\pi$ and $0.9\pi-\pi$, $4 < p_{T} < 6$ GeV/$c$, in $0-80 \%$ central collisions. The points are unlike-sign pairs with the $J/\psi$ signal, fitted by a Crystal Ball plus second order polynomial function. The blue solid line histogram shows the like-sign background. Bottom: $1/R$ weighted $J/\psi$ yield vs. $\phi$-$\Psi$ with fitted $v_2$. \label{fig1_yieldAndV2Fit}}
\end{figure}

The following method has been used to calculate the $v_2$ of $J/\psi$. Firstly, measurements of $\phi$-$\Psi$, ranging from 0 to $\pi$, were divided into 10 bins. Here $\phi$ is the azimuthal angle of the $J/\psi$ candidate, and $\Psi$ is the azimuthal angle of the event plane reconstructed from TPC tracks with the azimuthally nonuniform detector efficiency corrected for \cite{Poskanzer98}. The event plane resolution \cite{Poskanzer98} ($R$) is different for different centrality ranges, as listed in Table~\ref{tab:eventPlaneResolution}. Then two bins at supplementary angles were combined into one. For example, the bin at $0-0.1\pi$ is combined with $0.9\pi-\pi$, and the invariant mass distribution of electron/positron pairs in this combined $\phi$-$\Psi$ bin is shown in the top of Fig.~\ref{fig1_yieldAndV2Fit}. To avoid bias from different event plane resolution for different centrality, entries in the histogram were weighted by according $1/R$ \cite{Masui:2012zh}. The weighted $J/\psi$ yield within this combined $\phi$-$\Psi$ bin was obtained by fitting the $e^+e^-$ invariant mass distribution with a Crystal Ball function \cite{Gaiser:1982yw} signal on top of a second order polynomial background, as shown in the plot. The Crystal Ball function connects a Gaussian core with a power-law tail at low mass to account for daughter energy loss fluctuations and $J/\psi$ radiative decays. Then $v_{2}$ was obtained by fitting the weighted $J/\psi$ yield vs. $\phi$-$\Psi$ with a functional form of $A(1+2v_2\mathrm{cos}(2(\phi-\Psi)))$, as shown in the bottom of Fig.~\ref{fig1_yieldAndV2Fit}. Finally, the observed $v_2$ was scaled by $\langle1/R\rangle$ to obtain the true $v_2$ \cite{Masui:2012zh}.

\begin{table}[b]
\caption{\label{tab:eventPlaneResolution}%
Event plane resolution (R) for different centralities}
\begin{ruledtabular}
\begin{tabular}{l|r|r|r|r|r|r|r|r}
\textrm{cent (\%)}&
\textrm{ 0-10}&
\textrm{10-20}&
\textrm{20-30}&
\textrm{30-40}&
\textrm{40-50}&
\textrm{50-60}&
\textrm{60-70}&
\textrm{70-80}\\
\colrule
R & 0.600 & 0.748 & 0.805 & 0.787 & 0.719 & 0.608 & 0.478 & 0.364\\
\end{tabular}
\end{ruledtabular}
\end{table}

Three dominant sources of systematic error have been investigated for this measurement: assumptions in the $v_2$ calculation method, hadron contamination for the daughter $e^+e^-$ pairs, and the non-flow effect. The first source can be estimated from the difference in $v_2$ calculated by methods with different assumptions. Two other methods are used here. One is similar to the original method, except that the $J/\psi$ yield in each combined $\phi-\Psi$ bin was not obtained from fitting, but from subtracting the like-sign background from unlike-sign distribution within the possible invariant mass range of $J/\psi$ ($2.9-3.3$ GeV/$c^2$). In the other method, the overall $v_{2}$ of both signal and background was measured first as a function of invariant mass, and then it was fitted with an average of $J/\psi$ $v_{2}$ and background $v_{2}$ weighted by their respective yields vs. invariant mass \cite{Borghini:2004ra}. The systematic error from hadron contamination can be estimated from the difference in calculated $v_2$ with different electron/positron identification cuts. While the original cuts aim for the best $J/\psi$ significance, a purer electron/positron sample can be obtained from a set of tighter cuts. The overall systematic uncertainty for the first two sources was estimated from the maximum difference between the calculated $v_2$ with the $3\times2$ = 6 combinations of $v_2$ methods and electron/positron identification cut sets mentioned above. Besides elliptic flow, there are also some other two- and many-particle correlations due to, for example, resonance decay and jet production. When $v_2$ of a particle is measured, other particles having non-flow correlations with the measured particle are more likely to be azimuthally nearby, drawing the reconstructed event plane closer to the measured particle, and make the measured $v_2$ larger than its real value. To estimate this non-flow influence on the $v_2$ measurement, a method of scaling non-flow in $p+p$ collisions to that in Au + Au collisions \cite{Adams:2004wz} was employed. This method assumes that 1) $J/\psi$-hadron correlation in $p$ + $p$ collisions is entirely due to non-flow, and 2) the non-flow correlation to other particles per $J/\psi$ in Au + Au collisions is similar to that in $p$ + $p$ collisions. Under these assumptions, it can be deduced that the non-flow influence on measured $J/\psi$ $v_2$ in Au+Au collisions is $\langle\displaystyle\sum\limits_{i}\cos2(\phi_{J/\psi}-\phi_i)\rangle/M\overline{v_2}$. Here the sum is over all measured charged hadrons and the average is over $J/\psi$ in $p+p$ collisions. $M$ and $\overline{v_2}$ are the multiplicity and average elliptic flow of charged hadrons in Au+Au collisions, respectively. Since the away side correlation may be greatly modified by the medium in heavy-ion collisions, this procedure gives an upper limit of the non-flow effect. Detector acceptance and efficiency variation with $p_T$, centrality and rapidity may lead to a biased $J/\psi$ sample, which may induce some systematic effects when $v_2$ also changes with these parameters. But these effects are estimated to be negligible compared to statistical errors.

\begin{figure}
\includegraphics[width=1.\columnwidth]{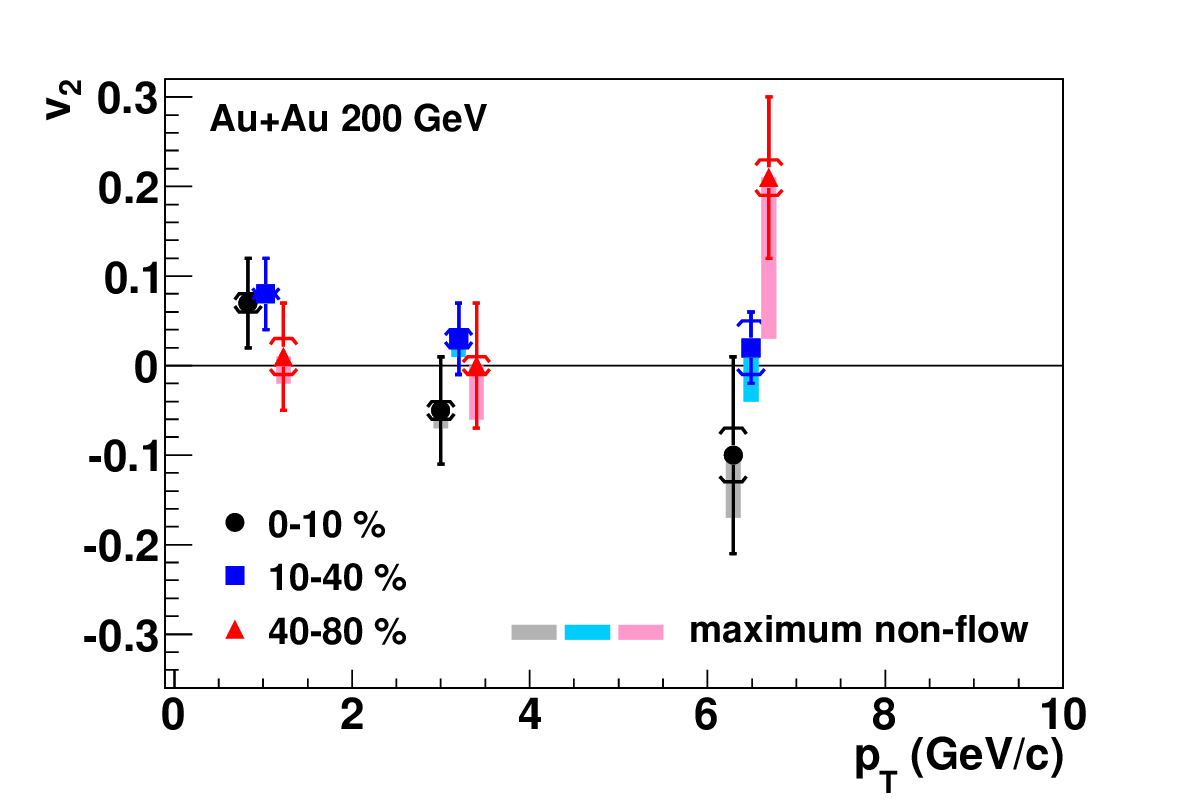}%
\caption{(color online) $v_{2}$ vs. $p_{T}$ for $J/\psi$ in different centrality bins. The brackets represent systematic errors estimated from differences between different methods and cuts. The boxes show the estimated maximum possible range of $v_2$ if the non-flow influence is corrected (see text). The $p_{T}$ bins for $J/\psi$ are $0-2$, $2-5$ and $5-10$ GeV/$c$. The mean $p_{T}$ in each bin for the $J/\psi$ sample used for $v_{2}$ calculation is drawn, but is shifted a little for some centralities so that all points can be seen clearly. \label{fig2_v2PtCent}}
\end{figure}

Figure~\ref{fig2_v2PtCent} shows $J/\psi$ $v_{2}$ as a function of transverse momentum for different centralities. Due to the non-flow effect, the real $v_2$ can be lower than the measured value shown in the plot. The boxes indicate the maximum magnitude of the non-flow influence. Data from the central trigger, minimum bias trigger and high tower triggers are used for the $0-10$ \% most central bin, while only minimum bias and high tower triggered events are used for other centrality bins. Considering errors and the magnitude of non-flow, $J/\psi$ $v_{2}$ is consistent with 0 for $p_{T} > 2$ GeV/$c$ for all measured centrality bins. Light particles usually have a larger $v_{2}$ in the intermediate centrality than in the most central and peripheral collisions. This can be explained by a larger initial spatial eccentricity in the intermediate centrality, which is transferred into final state momentum anisotropy due to different pressure gradients in different directions, when there are sufficient interactions in the medium. However, no strong centrality dependence for $J/\psi$ $v_2$ has been observed with the statistical significance of the data.

\begin{figure}
\includegraphics[width=1.\columnwidth]{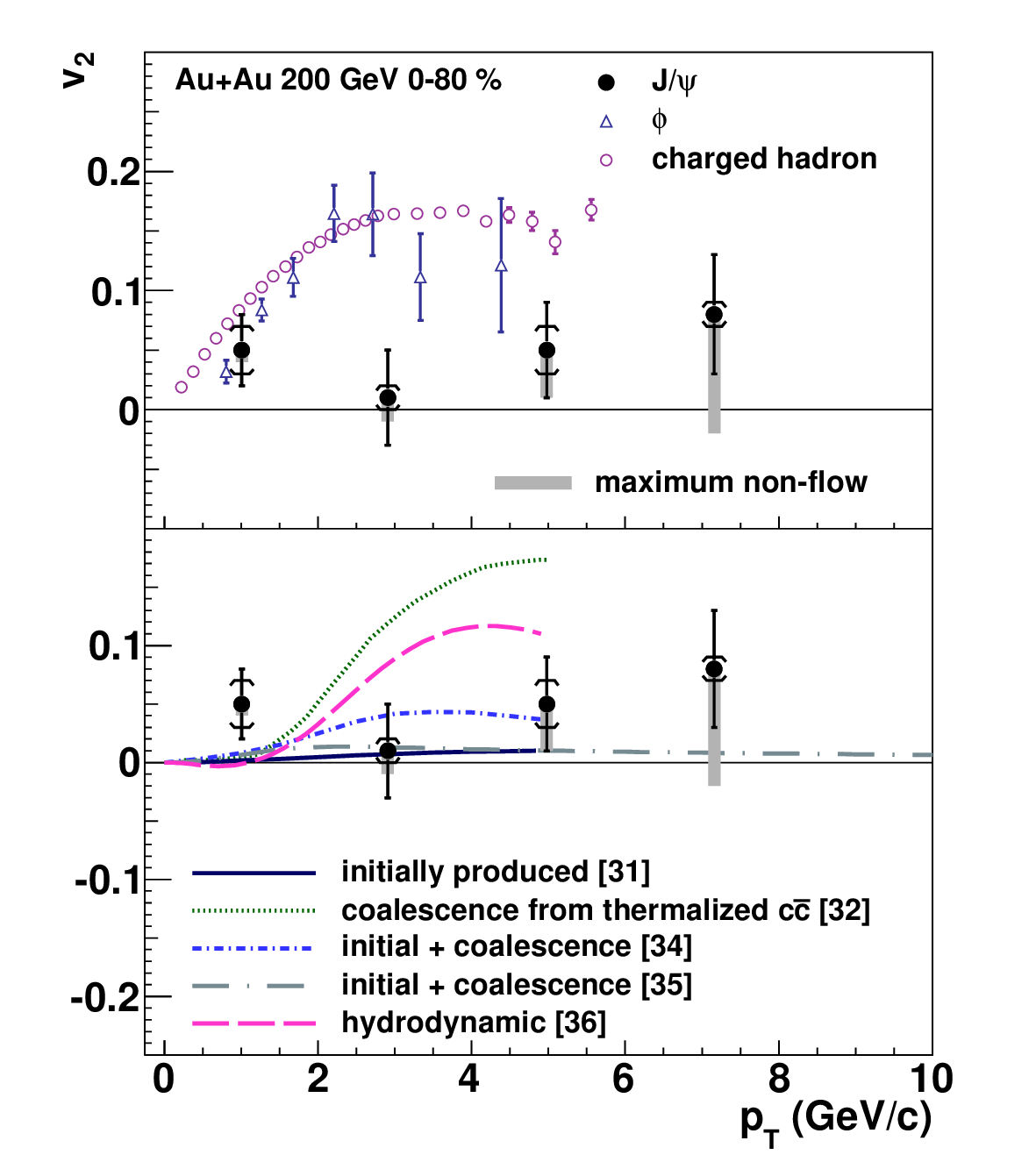}%
\caption{(color online) $v_{2}$ vs. $p_{T}$ for $J/\psi$ in $0-80$ \% central events comparing with charged hadrons \cite{Adams:2003zg} and the $\phi$ meson \cite{Abelev:2007rw} (upper panel) and theoretical calculations \cite{Yan:2006ve, Greco:2003vf, Ravagli:2007xx, Zhao:2008vu, Liu:2009gx, UllrichHeinz} (lower panel). The brackets represent systematic errors estimated from differences between different methods and cuts. The boxes show the estimated maximum possible range of $v_2$ if the non-flow influence is corrected. The $p_{T}$ bins for $J/\psi$ are $0-2$, $2-4$, $4-6$ and $6-10$ GeV/$c$, and the mean $p_{T}$ in each bin for the $J/\psi$ sample used for $v_{2}$ calculation is drawn. \label{fig3_v2PtCombined}}
\end{figure}

The top panel of Fig.~\ref{fig3_v2PtCombined} shows $J/\psi$ $v_{2}$ for $0-80$ \% central collisions as a function of transverse momentum. For reference, two other sets of $v_2$ measurements are also plotted, one is for charged hadrons (dominated by pions) \cite{Adams:2003zg} and the other is for the $\phi$ meson \cite{Abelev:2007rw} which is heavier than the pion but not as heavy as the $J/\psi$. Unlike $v_2$ of hadrons consisting of light quarks, $J/\psi$ $v_{2}$ at $p_{T} > 2$ GeV/$c$ is found to be consistent with zero within statistical errors. However, the significant mass difference between $J/\psi$ and light particles makes the direct comparison of $v_2$ vs. $p_T$ less conclusive. For example, for the same velocity at $y=0$, the $p_T$ of $J/\psi$ at 3.0 GeV/$c$ corresponds to $p_T$ of pions ($\phi$) at 0.14 (1.0) GeV/$c$. Thus comparisons between the experimental result and theoretical calculations are needed.

\begin{table}[b]
\caption{\label{tab:chi2P}%
Difference between model calculations and data. The p-value is the probability of observing a $\chi^{2}$ that exceeds the current measured $\chi^{2}$ by chance, even for a correct model. The estimated upper limit of non-flow effect is not included in this calculation.}
\begin{ruledtabular}
\begin{tabular}{lrr}
\textrm{theoretical calculation}&
\textrm{$\chi^{2}/\rm NDF$}&
\textrm{p-value}\\
\colrule
initially produced \cite{Yan:2006ve} & 2.6 / 3 & $4.6\times10^{-1}$\\
coalescence from thermalized $c\bar{c}$ \cite{Greco:2003vf}  & 16.2 / 3 & $1.0\times10^{-3}$\\ 
initial + coalescence \cite{Zhao:2008vu} & 2.0 / 3 & $5.8\times10^{-1}$\\ 
initial + coalescence \cite{Liu:2009gx} & 4.2 / 4 & $3.8\times10^{-1}$\\
hydrodynamic \cite{UllrichHeinz} & 7.0 / 3 & $7.2\times10^{-2}$\\ 
\end{tabular}
\end{ruledtabular}
\end{table}

In the bottom panel of Fig.~\ref{fig3_v2PtCombined}, a comparison is made between the measured $J/\psi$ $v_{2}$ and various theoretical calculations, and a quantitative level of difference is shown  in Table~\ref{tab:chi2P} by $\chi^2/\rm NDF$ and the p-value. $v_{2}$ of $J/\psi$ produced by initial pQCD processes is predicted to stay close to zero \cite{Yan:2006ve}. Although anomalous suppression in the hot medium due to color screening are considered in the model, the azimuthally different suppression along the different path lengths in azimuth leads to a limited $v_2$ beyond the sensitivity of the current measurement. On the contrary, if charm quarks get fully thermalized and $J/\psi$ are produced by coalescence from the thermalized flowing charm quarks at the freeze-out, the $v_{2}$ of $J/\psi$ is predicted to reach almost the same maximum magnitude as $v_2$ of light flavor mesons, although at a larger $p_T$ (around 4 GeV/$c$) due to the significantly larger mass of $J/\psi$ \cite{Greco:2003vf}. This is nearly 3$\sigma$ above the measurement for $p_{T} > 2$ GeV/$c$, leading to a large $\chi^{2}/\rm NDF$ of 16.2/3 and a small p-value of $1.0\times10^{-3}$, and is thus inconsistent with the data. Models that include $J/\psi$ from both initial production and coalescence production in the transport model \cite{Yan:2006ve,Ravagli:2007xx} predict a much smaller $v_{2}$ \cite{Zhao:2008vu,Liu:2009gx}, and are consistent with our measurement. In these models, $J/\psi$ are formed continuously through the system evolution rather than at the freeze-out, so many $J/\psi$ could be formed from charm quarks whose $v_2$ has still not fully developed. Furthermore, the initial production of $J/\psi$ with very limited $v_2$ dominates at high $p_{T}$, thus the overall $J/\psi$ $v_{2}$ does not rise rapidly as for light hadrons. This kind of model also describes the measured $J/\psi$ nuclear modification factor over a wide range of $p_{T}$ and centrality \cite{Zebo:2012}. The hydrodynamic model, which assumes local thermal equilibrium, can be tuned to describe $v_2$ for light hadrons, but it predicts a $J/\psi$ $v_2$ that rises strongly with $p_T$ in the region $p_T<4$ GeV/$c$, and thus fails to describe the main feature of the data \cite{UllrichHeinz}. For heavy particles such as $J/\psi$, hydrodynamic predictions suffer from large uncertainties related to viscous corrections ($\delta f$) at freeze-out and the assumed freeze-out time or temperature. 

In summary, $J/\psi$ elliptic flow is presented as a function of transverse momentum for different centralities in $\sqrt{s_{NN}}$ = 200 GeV Au+Au collisions. Unlike light flavor hadrons, $J/\psi$ $v_{2}$ at $p_{T} > 2$ GeV/$c$ is consistent with zero within statistical errors. Comparing to model calculations, the measured $J/\psi$ $v_2$ values disfavor the scenario that $J/\psi$ with $p_{T} > 2$ GeV/$c$ are produced dominantly by coalescence from (anti-)charm quarks which are thermalized and flow with the medium.


\begin{acknowledgments}
We thank the RHIC Operations Group and RCF at BNL, the NERSC Center at LBNL and the Open Science Grid consortium for providing resources and support. This work was supported in part by the Offices of NP and HEP within the U.S. DOE Office of Science, the U.S. NSF, the Sloan Foundation, CNRS/IN2P3, FAPESP CNPq of Brazil, Ministry of Ed. and Sci. of the Russian Federation, NNSFC, CAS, MoST, and MoE of China, GA and MSMT of the Czech Republic, FOM and NWO of the Netherlands, DAE, DST, and CSIR of India, Polish Ministry of Sci. and Higher Ed., National Research Foundation (NRF-2012004024), Ministry of Sci., Ed. and Sports of the Rep. of Croatia, and RosAtom of Russia.
\end{acknowledgments}

\bibliography{Jpsi_v2}

\end{document}